\documentclass[conference]{IEEEtran}
\IEEEoverridecommandlockouts
\usepackage{cite}
\usepackage{xspace}
\usepackage{nicefrac}
\usepackage{amsmath,amssymb,amsfonts}
\usepackage{algorithmic}
\usepackage{graphicx}
\usepackage{subcaption}
\usepackage{booktabs}
\usepackage{textcomp}
\usepackage{xcolor}
\usepackage{colortbl}
\usepackage{multirow}
\usepackage{url}
\def\BibTeX{{\rm B\kern-.05em{\sc i\kern-.025em b}\kern-.08em
    T\kern-.1667em\lower.7ex\hbox{E}\kern-.125emX}}
\RequirePackage{letltxmacro}

\begin{document}

\title{Evaluating quality metrics through the lenses of psychophysical measurements of low-level vision
}

\author{
\IEEEauthorblockN{
Dounia Hammou\IEEEauthorrefmark{1},
Yancheng Cai\IEEEauthorrefmark{1},
Pavan Madhusudanarao\IEEEauthorrefmark{2},
Christos G. Bampis\IEEEauthorrefmark{2},
Zhi Li\IEEEauthorrefmark{2},
Rafał K. Mantiuk\IEEEauthorrefmark{1}
}
\IEEEauthorblockA{
\IEEEauthorrefmark{1}\textit{University of Cambridge}, Cambridge, UK, 
\{dh706, yc613, rafal.mantiuk\}@cl.cam.ac.uk
}
\IEEEauthorblockA{
\IEEEauthorrefmark{2}\textit{Netflix, Inc.}, Los Gatos, CA, USA, 
\{pmadhusudanarao, christosb, zli\}@netflix.com
}
}

% \author{
% \IEEEauthorblockN{
% Dounia Hammou,
% Yancheng Cai,
% Pavan Madhusudanarao,
% Christos G. Bampis,
% Zhi Li,
% Rafał K. Mantiuk
% }
% }

% \author{}

\newcommand{\figref}[1]{Figure~\ref{fig:#1}}
\newcommand{\secref}[1]{Section~\ref{sec:#1}}
\newcommand{\algoref}[1]{Algorithm~\ref{algo:#1}}
\newcommand{\chapterref}[1]{Chapter~\ref{chapter:#1}}
\newcommand{\appref}[1]{Appendix~\ref{app:#1}}
\newcommand{\tableref}[1]{Table~\ref{tab:#1}}
\newcommand{\etal}{et al.\xspace}
\newcommand{\degpers}{\,\nicefrac{deg}{s}\xspace}
\newcommand{\degperssq}{\,\nicefrac{deg}{s\textsuperscript{2}}\xspace}
\newcommand{\degree}{$^{\circ}$\xspace}
\newcommand{\Hz}{\,Hz\xspace}
\newcommand{\cdms}{\,cd/m$^2$\xspace}
\newcommand{\ppd}{\,ppd\xspace}
\newcommand{\lux}{\,lux\xspace}
\newcommand{\fourier}{\mathfrak{F}}
\newcommand{\infourier}{^{\mathfrak{F}}}

\LetLtxMacro{\originaleqref}{\eqref}
\renewcommand{\eqref}[1]{Eq.~\originaleqref{eq:#1}}

\newcommand{\suppsecref}[1]{Sec.~\ref{suppsec:#1} of the supplementary document}

\newcommand{\RM}[1]{\textcolor{brown}{\textnormal{(Rafal) #1}}}
\newcommand{\dounia}[1]{\textcolor{blue}{\textnormal{(Dounia) #1}}}
\newcommand{\CB}[1]{\textcolor{green}{\textnormal{(Christos) #1}}}
\newcommand{\ZL}[1]{\textcolor{purple}{\textnormal{(Zhi Li) #1}}}

\newcommand{\ourmethod}{Name-of-your-method}
\newcommand{\ourdataset}{Name-of-your-dataset}

\newcommand{\supplementaryHTML}{supplementary HTML report}

\newcommand{\code}[1]{\texttt{#1}}

% Use for subscript/index that is NOT a variable (index)
\newcommand{\ind}[1]{\text{#1}}

% \IEEEpeerreviewmaketitle

\maketitle 

\begin{abstract}
% Image and video quality metrics, such as SSIM, LPIPS, and VMAF, are aimed to predict the perceived quality of the evaluated content and are often claimed to follow the human visual system. Yet, few metrics directly model human visual perception, and most rely on hand-crafted formulas or training datasets to achieve alignment with perceptual data. In this paper, we propose a set of tests for full-reference quality metrics that examine their ability to model several aspects of low-level human vision: contrast sensitivity, contrast masking, and contrast matching. The tests are meant to provide additional scrutiny for newly proposed metrics. We use our tests to analyze 34 existing image and video quality metrics and find their strengths and weaknesses, such as the ability of LPIPS and MS-SSIM to predict contrast masking and the poor performance of VMAF in this task. We further find that the popular SSIM metric overemphasizes differences in high spatial frequencies, but its multi-scale counterpart, MS-SSIM, addresses this shortcoming. Such findings cannot be easily made using existing evaluation protocols. 
Image and video quality metrics, such as SSIM, LPIPS, and VMAF, aim to predict perceived visual quality and are often assumed to reflect principles of human vision. However, relatively few metrics explicitly incorporate models of human perception, with most relying on hand-crafted formulas or data-driven training to approximate perceptual alignment. In this paper, we introduce a set of tests for full-reference quality metrics that evaluate their ability to capture key aspects of low-level human vision: contrast sensitivity, contrast masking, and contrast matching. These tests provide an additional framework for assessing both established and newly proposed metrics. We apply the tests to 34 existing quality metrics and highlight patterns in their behavior, including the ability of LPIPS and MS-SSIM to predict contrast masking and the tendency of SSIM to overemphasize high spatial frequencies, which is mitigated in MS-SSIM, and the general inability of metrics to model supra-threshold contrast constancy. Our results demonstrate how these tests can reveal properties of quality metrics that are not easily observed with standard evaluation protocols.

\end{abstract}

\begin{IEEEkeywords}
Image and video quality assessment, low-level vision, contrast sensitivity function (CSF), contrast detection and masking, contrast constancy
\end{IEEEkeywords}

\section{Introduction}
\label{sec:intro}

Objective image and video quality metrics, such as PSNR, SSIM \cite{wang2004image}, LPIPS \cite{zhang2018perceptual}, and VMAF \cite{li2018vmaf}, are widely used in quality of experience (QoE) research and standardization activities. They support the evaluation of image and video processing algorithms \cite{wang2020deep, hanji2022comparison}, the development and optimization of compression systems \cite{zhang2023survey}, and the large-scale assessment of visual quality where subjective testing is impractical. To improve perceptual relevance, many metrics are designed and validated based on their agreement with subjective quality measurements, including mean opinion scores (MOS) \cite{ITU-T_P910} and just-noticeable difference (JND) or just-objectionable difference (JOD) scales \cite{Perez_Ortiz_2020}. Hence, the primary methodology for evaluating objective quality metrics is to measure their correlation with subjective quality scores. While this approach remains essential, it is influenced by factors such as observer variability, experimental noise \cite{Krasula_2016,Ragano_2025}, and differences in content, distortion types, and test protocols across datasets \cite{Perez_Ortiz_2020}. As a result, metrics may exhibit varying performance across datasets, and correlation-based results alone provide limited insight into the characteristics that drive metric behavior.

At the same time, the early stages of the human visual system have been extensively studied using psychophysical experiments, leading to well-established measurements and models of visual sensitivity and discrimination. Such experiments rely on controlled stimuli, such as Gabor patterns, and yield reproducible results that characterize fundamental properties of human vision. These properties are known to contribute to higher-level perceptual judgments and, ultimately, perceived visual quality. 
In this paper, we present an analysis framework that uses psychophysically motivated stimuli to characterize the behavior of image and video quality metrics. The proposed framework complements conventional subjective evaluations by examining metric responses to controlled variations in spatial frequency, contrast magnitude, and temporal modulation. Metric outputs are compared with psychophysical data and established models describing contrast sensitivity \cite{ashraf2024castlecsf}, contrast masking \cite{foley1994human, gegenfurtner1992contrast}, temporal contrast sensitivity \cite{cai2024elatcsf}, and supra-threshold contrast perception \cite{georgeson1975contrast, switkes1999comparison}.
The resulting analysis provides a structured and interpretable characterization of metric behavior, highlighting similarities and differences between metric responses and known properties of human visual perception. This characterization is particularly relevant for modern learning-based metrics, whose internal representations are difficult to interpret but whose behavior can be systematically probed using controlled stimuli. The proposed approach does not aim to replace subjective testing, but rather to support metric development, comparison, and application by offering additional insight into the low-level characteristics captured by objective quality metrics. 
An open-source evaluation framework will be released after publication. 

% to enable the community to evaluate their metrics using the proposed methodology. Additionally, a webpage reporting the performance of all evaluated metrics will be maintained and updated regularly as new metrics are added. %An overview of the supplementary webpage can be found in: ... \dounia{I will include the webpage with all metric results as a ZIP file in google drive.}

% The main contributions of this work are:
% \begin{itemize}
% \item A set of 11 psychophysically motivated tests for analyzing the response of image and video quality metrics to established characteristics of low-level human vision.
% \item A comparative analysis of 33 image and video quality metrics, providing insight into their perceptual behavior and alignment with psychophysical measurements.
% \end{itemize}

% \begin{figure}[]
% \centering
% \includegraphics[width=0.9\linewidth]{figs/teaser_figure.pdf}
% \caption{Can a quality metric model the low-level human vision? In this work, we evaluate how well existing image and video quality metrics model the characteristics of low-level human visions, such as contrast detection and masking. We assess their performance by measuring their alignment with well-established psychophysical data or models.}
% \label{fig:testing_methodology}
% \end{figure}

\section{Related work}
\label{sec:related_works}

% \dounia{I will include MAD as well \cite{wang2008maximum}}

The dominant methodology for testing image and video quality metrics is to compare the predictions with subjective data in terms of correlation coefficients (Pearson's, Spearman's, or Kendall's coefficients) or root-mean-square-error (RMSE). The main weakness of such an approach is that it does not account for the measurement error associated with the subjective data (MOS, DMOS, JODs). Krasula et al. \cite{Krasula_2016} addressed this shortcoming by posing the metrics' task as a classification problem. Their method tests whether a metric can classify condition pairs into those that are of similar and different quality and then, for those that are of different quality, whether it can predict which one has a higher quality. More recently, Ragano et al. \cite{Ragano_2025} proposed the Constrained Concordance Index (CCI), which measures the metric's capability to accurately rank pairs where MOS has high precision and ignores the ones with uncertain MOS. All these measures help to better discriminate between metrics performance on a given dataset, but they provide little insight into why metrics perform poorly or well. Our tests are not meant to replace metric performance measures but to complement them by illustrating the strengths and weaknesses of the metrics. 

% Psychophysical stimuli were used before to calibrate and test visual difference predictors: HDR-VDP-2 \cite{mantiuk2011hdr}, FovVideoVDP \cite{mantiuk2021fovvideovdp}, and ColorVideoVDP \cite{cvvdp}. However, those works involved a limited range of such stimuli and lacked a comprehensive comparison with other quality metrics. 
% Other studies proposed datasets to evaluate the models. For instance, the ModelFest dataset \cite{Watson_Ahumada_2005} offers a set of 43 contrast discrimination stimuli for testing visual models. Alam et al. \cite{Alam_2014} created a dataset for local masking in natural images for testing contrast masking models. We do not use either of these datasets as they provide less structured data that is more difficult to visualize and interpret. 

Psychophysical stimuli have previously been used to calibrate and evaluate visual difference predictors such as HDR-VDP-3 \cite{mantiuk2023hdr}, FovVideoVDP \cite{mantiuk2021fovvideovdp}, and ColorVideoVDP \cite{cvvdp}. However, these studies relied on a limited range of stimuli and did not provide a comprehensive comparison with other quality metrics.
Other studies proposed datasets for testing psychophysical models. For example, the ModelFest dataset \cite{Watson_Ahumada_2005} provides 43 contrast discrimination stimuli for evaluating visual models, while Alam et al. \cite{Alam_2014} created a dataset for studying local masking in natural images. We do not use either dataset, as their data are less structured and more difficult to visualize and interpret. 
Furthermore, several works have explored the presence of the contrast sensitivity function (CSF) in neural networks \cite{tariq2020deep,Akbarinia_2023}. For example, Tariq \etal{} \cite{tariq2020deep} found that CNN features aligned with the CSF show higher correlation with human quality judgments. Similarly, Akbarinia \etal{} \cite{Akbarinia_2023} demonstrated that networks trained on natural visual tasks, such as object recognition or depth estimation, match the CSF as they focus on visual information similar to what humans use, allowing the networks to capture the limitations of the human visual system. However, contrast sensitivity alone primarily explains human performance only for very low-contrast stimuli on uniform backgrounds. Therefore, in this work, we investigate both near-threshold and supra-threshold aspects of human vision, including contrast masking and contrast matching.

More recently, Cai et al. \cite{Cai_2025} used similar methods to ours to investigate whether computer vision foundation models learned low-level human vision features. Here, we focus on image and video quality metrics and customize our tests for such. For example, we include temporal (flicker) and color matching tests that are relevant for video and color metrics. 

\section{Proposed evaluation framework}
\label{sec:methodology}

\begin{figure*}[!htb]
\centering
\includegraphics[width=\linewidth]{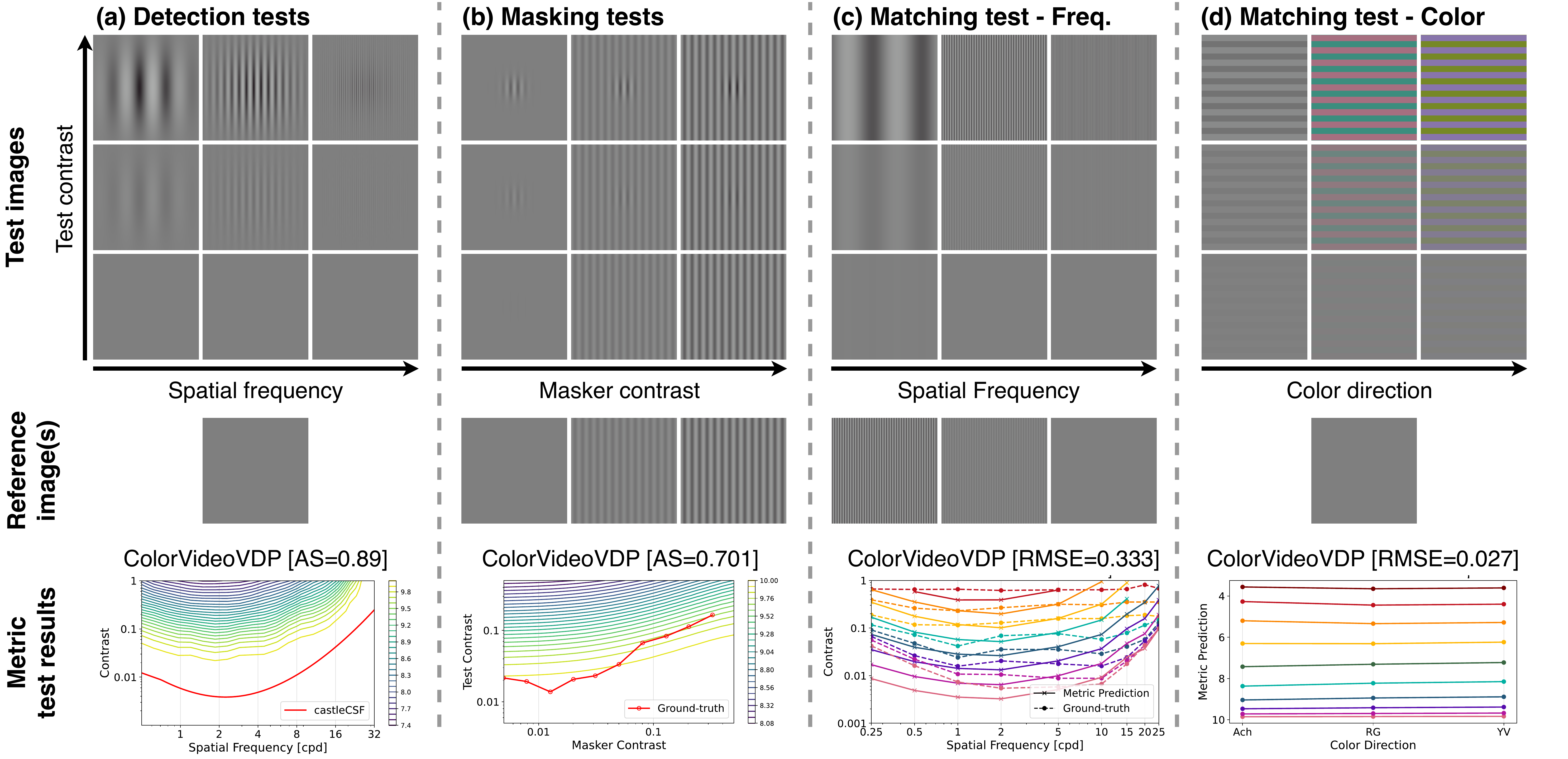}
\caption{A representation of the methodologies employed to evaluate the metrics on the various tests. The first row represents a grid of the test images. The second row represents the reference image(s) for each test. For tests where the reference image was a uniform field, we showcase only one image. The third row showcases the metric results on the specific test, for example, a contour plot for the detection test, as well as the performance score (an alignment score (AS) or RMSE).}
\label{fig:testing_methodology}
\end{figure*}

Our testing methodology mimics classical psychophysical experiments used to study low-level human vision. Depending on the perceptual task --- such as detection or matching --- these experiments adopt different measurement procedures. In this work, we evaluate quality metrics using methodologies designed to reflect these experimental procedures.

Psychophysical experiments require stimuli calibrated in physical units --- pixel values as luminance (\cdms{}) or CIE LMS cone responses, and image size in visual degrees. Some metrics (ColorVideoVDP, HDR-VDP, FLIP) let us pass the stimuli in physical units. However, most of the quality metrics ignore the physical image resolution and expect display-encoded images in the sRGB color space as input. For that reason, we had to make a few assumptions and restrict the range of physical parameters. The physical resolution was assumed to be near 60 pixels per degree (ppd), which corresponds to the ITU-T BT.2100 recommendation for the FullHD television (3 display heights).
We varied this resolution in some tests to better reproduce high-frequency patterns. The display peak luminance was assumed to be 100\cdms{} (typical for desktop displays), and the mean luminance for most patterns was 21.4\cdms{}. The patterns were generated in a linear color space (luminance or CIE 2006 LMS cone responses) and then converted into the sRGB color space. Furthermore, we ensured that all images were passed to the metric either as floating-point numbers or integers quantized to 16 bits to ensure minimal loss of information.

Our tests can be grouped into two categories, each targeting a different aspect of early visual processing.%, which we discuss next. 

\subsection{Detection tests}

In a detection experiment, a participant is shown two intervals --- one with a detection pattern at a given contrast and one without it --- and then asked to determine in which interval the pattern was present. The outcome of such a measurement is a psychometric function describing the probability of detection as a function of contrast. 
We adopt an analogous procedure for quality metrics. The quality metric plays the role of the observer: a test image containing a stimulus is compared against a reference image without a stimulus (for example a uniform field in contrast detection tests, see (a) in \figref{testing_methodology}). By varying stimulus contrast and for example spatial frequency, the metric produces a set of responses that can be interpreted as a 2D detection surface. We visualize this surface as a contour plot (bottom of \figref{testing_methodology}-(a) and (b)). These contours can then be directly compared to psychophysical data or models, such as a contrast sensitivity function (CSF), shown as the red solid line in \figref{testing_methodology}-(a). A metric that correctly models contrast detection, for instance, should exhibit contour lines aligned with the CSF.

\paragraph{Contrast detection test}

In this test, the reference image was a uniform field, and the test image contained a Gabor patch (see \figref{testing_methodology}-(a)) of varying spatial frequencies, luminance or area sizes (x-axis) and contrast levels (y-axis).
The resolution of the stimuli was $1920 \times 1080$. We used a recent contrast sensitivity model, castleCSF \cite{ashraf2024castlecsf}, to predict the detection threshold of the human observer. The model has been shown to predict multiple psychophysical datasets and provide predictions for an average observer.

\paragraph{Contrast masking tests}
In these tests, the reference image contained a masker, such as a sinusoidal grating (phase coherent masking) or a broadband noise (phase incoherent masking) of varying contrast levels (x-axis). The test images contained the same masker plus a Gabor patch of varying contrast levels (y-axis) --- see column (b) in \figref{testing_methodology}. The red lines represent user data collected in \cite{foley1994human} for sinusoidal maskers, and in \cite{gegenfurtner1992contrast} for broadband noise.

\paragraph{Flicker detection test}

In this test, the reference video was a uniform field, and the test video contained a disk modulated over time (sinusoidal modulation) of varying temporal frequencies (x-axis) and contrast levels (y-axis). 
The resolution of the stimuli was $256 \times 256$ for computational reasons, its framerate was 120\,fps, and its duration was 1\,s. We used a recent temporal contrast sensitivity model, elaTCSF \cite{cai2024elatcsf}, to predict the detection thresholds. 

\subsection{Contrast matching tests}
In the contrast matching experiment, an observer is asked to adjust the contrast of a test stimulus to match the perceived magnitude of contrast in a reference stimulus, across spatial frequencies (see column (c) in \figref{testing_methodology}) or the modulation direction in a color space (e.g., the achromatic pattern is matched to the red-green pattern, see column (d) in \figref{testing_methodology}). 

\paragraph{Contrast matching across spatial frequencies test}

To mimic an observer in this task, we use the quality metric to identify the test contrast $c_t$ that matches a reference contrast $c_r$ at a reference $\rho_r$ across different spatial frequencies $\rho_t$. Specifically, we want a metric prediction between a uniform field $U$ and a sinusoidal test grating $S\left ( \rho_t, c_t \right )$ of frequency $\rho_t$ and contrast $c_t$ to be the same as the metric prediction for the uniform field and a reference sinusoidal grating, $S\left ( \rho_r, c_r \right )$:
\begin{equation}
\label{eq:matching}
Q\left ( S\left ( \rho_t, c_t \right ), U \right ) = Q\left ( S\left ( \rho_r, c_r \right ), U \right )
\end{equation}
where $Q(.)$ is the metric prediction. Hence, for each reference contrast $c_r$ and test frequency $\rho_t$, we solved \eqref{matching} to find the test contrast $c_t$. The human data we use for comparison comes from the study of Georgeson and Sullivan \cite{georgeson1975contrast}, in which the reference frequency was at $\rho_r=5$\,cpd (cycles per degree), test frequencies varied between 0.25 and 25\,cpd, the background luminance was 10\cdms{}, and the effective resolution was 50\,ppd. The resolution of the images was $256 \times 256$. Contrast matching data is visualized as lines of matching contrast, i.e., the lines that indicate contrast that appears the same across frequencies, as shown at the bottom of column (c) of \figref{testing_methodology}. 

\paragraph{Contrast matching across color direction test}
% For this test, we used $256 \times 256$ square-wave patterns as test images, together with the matching data from Switkes et al. \cite{switkes1999comparison} (see column (d) in \figref{testing_methodology}). In this test, we evaluate whether the metric’s prediction for one color direction (e.g., red–green) aligns with that for another direction (e.g., achromatic) across contrast levels when compared to a uniform field (reference image). The metric results are visualized as a line connecting responses across color directions, as shown at the bottom of column (d) in \figref{testing_methodology}; ideally, this line should remain horizontal across directions.
In this test, we used the matching data from Switkes et al. \cite{switkes1999comparison}, who matched contrast along different contrast modulation directions in a color space. The data allowed us to define the contrasts in the achromatic (grayscale), red-green, and yellow-violet color directions so that the perceived magnitude of contrast for each is matched, as shown in column (d) of \figref{testing_methodology}. 
To evaluate the quality metrics, we generated test images of a resolution of $256 \times 256$ with square-wave patterns in the different color directions so that the perceived magnitude of contrast for each is matched when compared to a uniform field (reference image). We plot the metrics' response for each such triplet, connected by a line, as shown at the bottom of column (d) in \figref{testing_methodology};  ideally, if a metric correctly matches contrast across color directions, the response of the metric should be the same, and the lines should be horizontal.

\section{Aggregate performance measures}

In addition to the plots used to visualize metric performance (see the bottom of \figref{testing_methodology}), we compute two summary measures: alignment score and RMSE \cite{cai2024elatcsf}. The alignment score, used for detection tests, quantifies how well a metric’s predictions follow the contrast threshold. Specifically, we evaluate predictions at multiples of the contrast threshold; strong performance is indicated by high correlation between the metric’s predictions and these multipliers. RMSE, used for contrast matching, measures prediction accuracy relative to ground truth. For the spatial-frequency test, it captures the difference between the predicted and true test contrasts. For the color-direction test, it measures the normalized difference in quality scores across color directions. 

It should be noted that such measures lack the interpretation and detail provided by the plots but give a high level overview and facilitate comparison across multiple metrics. 

% \tableref{metrics_results} shows the alignment scores (the higher, the better) for contrast detection and masking and RMSE (the lower, the better) for contrast matching. As expected, the visual difference predictors that explicitly model low-level vision show the best alignment with the human data. LPIPS variants and DISTS stand out as those that model contrast masking well. MS-SWD models the detection of both chromatic and achromatic patterns well. Only FovVideoVDP and ColorVideoVDP model flicker. The RMSE values for contrast matching do not explain the performance well as they indicate that the metrics without spatial processing (e.g., PSNR-Y, CIEDE2000) perform the best. This is because those metrics have flat contrast response across all frequencies, which happen to capture contrast constancy at high contrast values --- see row (h) of \figref{metrics_results_main_fig}.

\begin{figure*}[!htb]
\centering
\includegraphics[width=\linewidth]{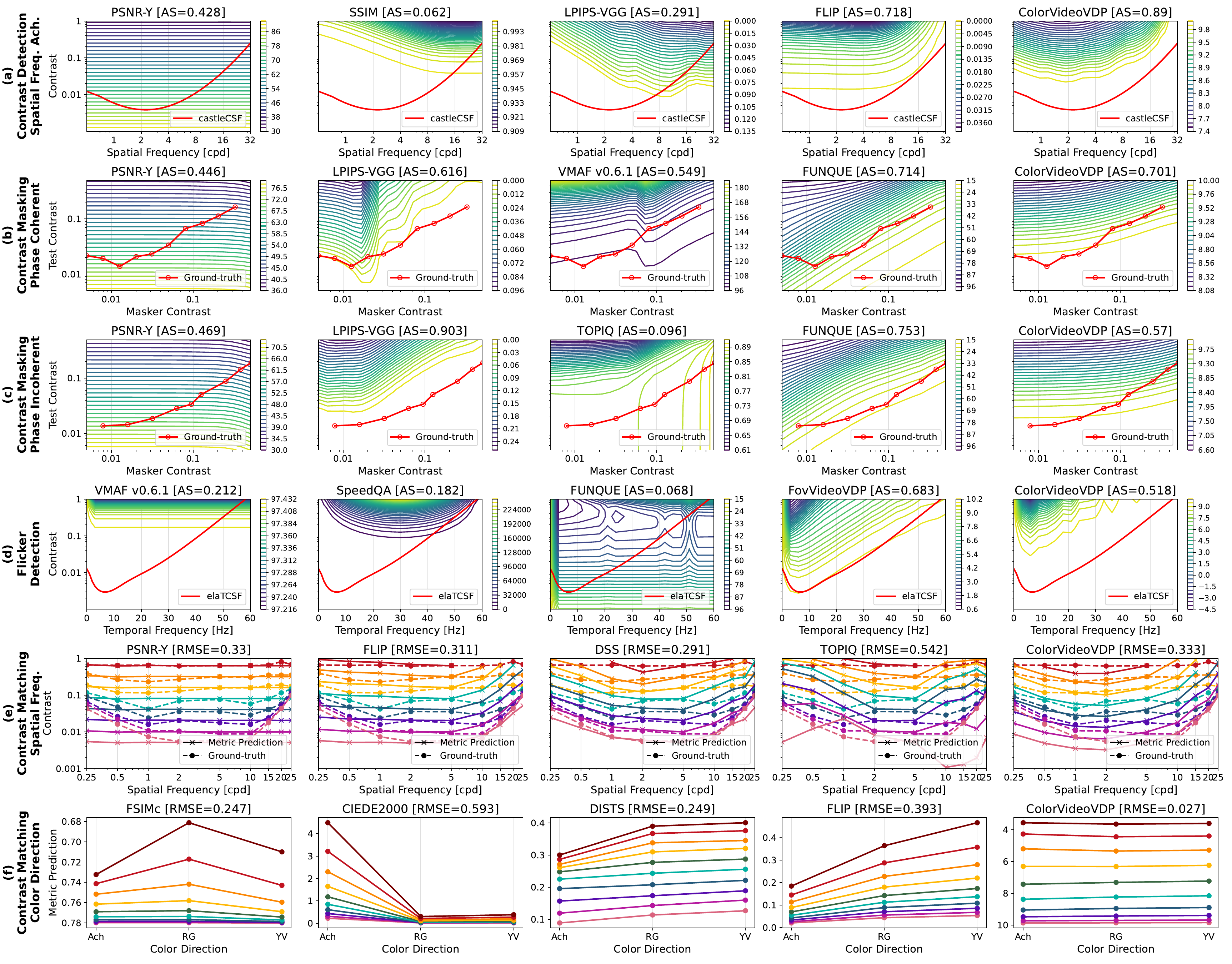}
\caption{Metric predictions compared to the human data for selected quality metrics. Each row represents a test, and each column corresponds to a different metric. The detection tests are in rows (a)--(d), and the alignment score ($\uparrow$) for each is reported in the square brackets. The red lines show the human performance based on castleCSF for (a), measurements for (b)--(c), and elaTCSF for (d). The contract matching tests are reported in rows (e) and (f) with the RMSE ($\downarrow$) in the square brackets. The human performance is shown as dashed lines in (e). The perfect alignment for matching across color directions (f) should result in horizontal lines.}
\label{fig:metrics_results_main_fig}
\end{figure*}

\section{Experimental results and analysis}
\label{sec:experimental_results}

We evaluated 34 full-reference image and video quality metrics on our psychophysical tests. Only full-reference metrics were included, as the tests require a comparison to a reference. %The metrics were selected to represent different approaches to quality assessment: metrics based on traditional features, deep-learning, or psychophysical models; image and video metrics; luminance and color metrics. The tested metrics are listed in \tableref{metrics_results}.
For fairness, color tests are evaluated only on color metrics and temporal tests only on video metrics.

Given the large number of tests and metrics, we focus on the most informative tests in our analysis. We report metrics' performance in \tableref{metrics_results} and show plots for only the five most representative metrics in \figref{metrics_results_main_fig}. Readers are encouraged to consult the \supplementaryHTML{}\footnote{The HTML report can be found anonymously in: \url{https://drive.google.com/file/d/1SaN5yKb6Ff_y6S0AL0eQy2kFhxoo10eu/view?usp=sharing}} for complete results across all metrics.

\begin{table}[]
\centering
\caption{Alignment scores $\uparrow$ (contrast detection, masking and flicker) and RMSE $\downarrow$ (contrast matching) for all tested metrics. The colored background denotes the quantile across all the tested metrics: green --- top 25\%, yellow --- 25-50\%, orange 50-75\%, red --- bottom 25\%. An empty cell indicates that the metric does not offer color or temporal processing or the score could not be computed. (*) The metric returned identical values, and correlation could not be computed.}
\label{tab:metrics_results}
\resizebox{\linewidth}{!}{%
\begin{tabular}{|ll|c|cc|c|cc|}
\hline
\multicolumn{2}{|l|}{\multirow{2}{*}{Metric}} & \multirow{2}{*}{\begin{tabular}[c]{@{}c@{}}Contrast \\ Detection\end{tabular}} & \multicolumn{2}{c|}{Contrast Masking} & \multirow{2}{*}{\begin{tabular}[c]{@{}c@{}}Flicker \\ Detection\end{tabular}} & \multicolumn{2}{c|}{Contrast Matching} \\ 
 \cline{4-5} \cline{7-8} & &  & \multicolumn{1}{c|}{\begin{tabular}[c]{@{}c@{}} Phase\\Coherent\end{tabular}} & \multicolumn{1}{c|}{\begin{tabular}[c]{@{}c@{}} Phase\\Incoherent\end{tabular}} &  & \multicolumn{1}{c|}{\begin{tabular}[c]{@{}c@{}} Spatial\\Freq.\end{tabular}} & \multicolumn{1}{c|}{\begin{tabular}[c]{@{}c@{}} Color\\Direction\end{tabular}} \\ 
\hline \hline
\multicolumn{1}{|l|}{\multirow{8}{*}{ \rotatebox[origin=c]{90}{ \begin{tabular}[c]{@{}c@{}}Traditional metrics \end{tabular}}}} &PSNR-Y & \multicolumn{1}{c|}{\cellcolor[HTML]{fff8d6} 0.428} & \multicolumn{1}{c|}{\cellcolor[HTML]{ffd8b8} 0.446} & \multicolumn{1}{c|}{\cellcolor[HTML]{ffd8b8} 0.469} & \multicolumn{1}{c|}{\cellcolor[HTML]{f1f1f1} -} & \multicolumn{1}{c|}{\cellcolor[HTML]{cbe1cd} 0.33} & \multicolumn{1}{c|}{\cellcolor[HTML]{f1f1f1} -} \\ 
\cline{2-8} \multicolumn{1}{|l|}{} &SSIM \cite{wang2004image} & \multicolumn{1}{c|}{\cellcolor[HTML]{fdddd8} 0.062} & \multicolumn{1}{c|}{\cellcolor[HTML]{ffd8b8} 0.483} & \multicolumn{1}{c|}{\cellcolor[HTML]{fff8d6} 0.569} & \multicolumn{1}{c|}{\cellcolor[HTML]{f1f1f1} -} & \multicolumn{1}{c|}{\cellcolor[HTML]{ffd8b8} 0.507} & \multicolumn{1}{c|}{\cellcolor[HTML]{f1f1f1} -} \\ 
\cline{2-8} \multicolumn{1}{|l|}{} &MS-SSIM \cite{wang2003multiscale} & \multicolumn{1}{c|}{\cellcolor[HTML]{cbe1cd} 0.682} & \multicolumn{1}{c|}{\cellcolor[HTML]{fff8d6} 0.559} & \multicolumn{1}{c|}{\cellcolor[HTML]{fff8d6} 0.535} & \multicolumn{1}{c|}{\cellcolor[HTML]{f1f1f1} -} & \multicolumn{1}{c|}{\cellcolor[HTML]{cbe1cd} 0.23} & \multicolumn{1}{c|}{\cellcolor[HTML]{f1f1f1} -} \\ 
\cline{2-8} \multicolumn{1}{|l|}{} &GMSD \cite{xue2013gradient} & \multicolumn{1}{c|}{\cellcolor[HTML]{ffd8b8} 0.386} & \multicolumn{1}{c|}{\cellcolor[HTML]{fff8d6} 0.526} & \multicolumn{1}{c|}{\cellcolor[HTML]{fff8d6} 0.554} & \multicolumn{1}{c|}{\cellcolor[HTML]{f1f1f1} -} & \multicolumn{1}{c|}{\cellcolor[HTML]{ffd8b8} 0.459} & \multicolumn{1}{c|}{\cellcolor[HTML]{f1f1f1} -} \\ 
\cline{2-8} \multicolumn{1}{|l|}{} &MS-GMSD \cite{zhang2017gradient} & \multicolumn{1}{c|}{\cellcolor[HTML]{fff8d6} 0.407} & \multicolumn{1}{c|}{\cellcolor[HTML]{fff8d6} 0.571} & \multicolumn{1}{c|}{\cellcolor[HTML]{fff8d6} 0.528} & \multicolumn{1}{c|}{\cellcolor[HTML]{f1f1f1} -} & \multicolumn{1}{c|}{\cellcolor[HTML]{fff8d6} 0.356} & \multicolumn{1}{c|}{\cellcolor[HTML]{f1f1f1} -} \\ 
\cline{2-8} \multicolumn{1}{|l|}{} &VIFp \cite{sheikh2006image} & \multicolumn{1}{c|}{\cellcolor[HTML]{fdddd8} 0.019} & \multicolumn{1}{c|}{\cellcolor[HTML]{fdddd8} 0.297} & \multicolumn{1}{c|}{\cellcolor[HTML]{fdddd8} 0.38} & \multicolumn{1}{c|}{\cellcolor[HTML]{f1f1f1} -} & \multicolumn{1}{c|}{\cellcolor[HTML]{ffd8b8} 0.586} & \multicolumn{1}{c|}{\cellcolor[HTML]{f1f1f1} -} \\ 
\cline{2-8} \multicolumn{1}{|l|}{} &DSS \cite{balanov2015image} & \multicolumn{1}{c|}{\cellcolor[HTML]{fff8d6} 0.489} & \multicolumn{1}{c|}{\cellcolor[HTML]{cbe1cd} 0.746} & \multicolumn{1}{c|}{\cellcolor[HTML]{cbe1cd} 0.614} & \multicolumn{1}{c|}{\cellcolor[HTML]{f1f1f1} -} & \multicolumn{1}{c|}{\cellcolor[HTML]{cbe1cd} 0.291} & \multicolumn{1}{c|}{\cellcolor[HTML]{f1f1f1} -} \\ 
\cline{2-8} \multicolumn{1}{|l|}{} &NLPD \cite{laparra2016perceptual} & \multicolumn{1}{c|}{\cellcolor[HTML]{fff8d6} 0.484} & \multicolumn{1}{c|}{\cellcolor[HTML]{fff8d6} 0.531} & \multicolumn{1}{c|}{\cellcolor[HTML]{ffd8b8} 0.487} & \multicolumn{1}{c|}{\cellcolor[HTML]{f1f1f1} -} & \multicolumn{1}{c|}{\cellcolor[HTML]{cbe1cd} 0.279} & \multicolumn{1}{c|}{\cellcolor[HTML]{f1f1f1} -} \\ 
\hline \hline
\multicolumn{1}{|l|}{\multirow{6}{*}{ \rotatebox[origin=c]{90}{ \begin{tabular}[c]{@{}c@{}}Traditional\\color metrics \end{tabular}}}} &FSIMc \cite{zhang2011fsim} & \multicolumn{1}{c|}{\cellcolor[HTML]{fdddd8} 0.039} & \multicolumn{1}{c|}{\cellcolor[HTML]{fff8d6} 0.642} & \multicolumn{1}{c|}{\cellcolor[HTML]{cbe1cd} 0.633} & \multicolumn{1}{c|}{\cellcolor[HTML]{f1f1f1} -} & \multicolumn{1}{c|}{\cellcolor[HTML]{f1f1f1} -} & \multicolumn{1}{c|}{\cellcolor[HTML]{cbe1cd} 0.247} \\ 
\cline{2-8} \multicolumn{1}{|l|}{} &VSI \cite{zhang2014vsi} & \multicolumn{1}{c|}{\cellcolor[HTML]{fdddd8} 0.134} & \multicolumn{1}{c|}{\cellcolor[HTML]{fff8d6} 0.629} & \multicolumn{1}{c|}{\cellcolor[HTML]{fff8d6} 0.534} & \multicolumn{1}{c|}{\cellcolor[HTML]{f1f1f1} -} & \multicolumn{1}{c|}{\cellcolor[HTML]{f1f1f1} -} & \multicolumn{1}{c|}{\cellcolor[HTML]{ffd8b8} 0.458} \\ 
\cline{2-8} \multicolumn{1}{|l|}{} &MDSI \cite{nafchi2016mean} & \multicolumn{1}{c|}{\cellcolor[HTML]{fff8d6} 0.538} & \multicolumn{1}{c|}{\cellcolor[HTML]{ffd8b8} 0.484} & \multicolumn{1}{c|}{\cellcolor[HTML]{fff8d6} 0.53} & \multicolumn{1}{c|}{\cellcolor[HTML]{f1f1f1} -} & \multicolumn{1}{c|}{\cellcolor[HTML]{fdddd8} 0.602} & \multicolumn{1}{c|}{\cellcolor[HTML]{fff8d6} 0.337} \\ 
\cline{2-8} \multicolumn{1}{|l|}{} &HaarPSI \cite{reisenhofer2018haar} & \multicolumn{1}{c|}{\cellcolor[HTML]{fff8d6} 0.504} & \multicolumn{1}{c|}{\cellcolor[HTML]{fff8d6} 0.614} & \multicolumn{1}{c|}{\cellcolor[HTML]{ffd8b8} 0.484} & \multicolumn{1}{c|}{\cellcolor[HTML]{f1f1f1} -} & \multicolumn{1}{c|}{\cellcolor[HTML]{fff8d6} 0.385} & \multicolumn{1}{c|}{\cellcolor[HTML]{fff8d6} 0.344} \\ 
\cline{2-8} \multicolumn{1}{|l|}{} &sCIELab \cite{zhang1996spatial} & \multicolumn{1}{c|}{\cellcolor[HTML]{cbe1cd} 0.635} & \multicolumn{1}{c|}{\cellcolor[HTML]{ffd8b8} 0.447} & \multicolumn{1}{c|}{\cellcolor[HTML]{ffd8b8} 0.476} & \multicolumn{1}{c|}{\cellcolor[HTML]{f1f1f1} -} & \multicolumn{1}{c|}{\cellcolor[HTML]{fff8d6} 0.403} & \multicolumn{1}{c|}{\cellcolor[HTML]{ffd8b8} 0.45} \\ 
\cline{2-8} \multicolumn{1}{|l|}{} &FLIP \cite{andersson2020flip} & \multicolumn{1}{c|}{\cellcolor[HTML]{cbe1cd} 0.718} & \multicolumn{1}{c|}{\cellcolor[HTML]{ffd8b8} 0.449} & \multicolumn{1}{c|}{\cellcolor[HTML]{ffd8b8} 0.477} & \multicolumn{1}{c|}{\cellcolor[HTML]{f1f1f1} -} & \multicolumn{1}{c|}{\cellcolor[HTML]{cbe1cd} 0.311} & \multicolumn{1}{c|}{\cellcolor[HTML]{fff8d6} 0.393} \\ 
\hline \hline
\multicolumn{1}{|l|}{\multirow{4}{*}{ \rotatebox[origin=c]{90}{ \begin{tabular}[c]{@{}c@{}}Color\\difference\\measures \end{tabular}}}} &CIEDE2000 \cite{CIE_2018} & \multicolumn{1}{c|}{\cellcolor[HTML]{fff8d6} 0.428} & \multicolumn{1}{c|}{\cellcolor[HTML]{ffd8b8} 0.452} & \multicolumn{1}{c|}{\cellcolor[HTML]{ffd8b8} 0.474} & \multicolumn{1}{c|}{\cellcolor[HTML]{f1f1f1} -} & \multicolumn{1}{c|}{\cellcolor[HTML]{cbe1cd} 0.329} & \multicolumn{1}{c|}{\cellcolor[HTML]{fdddd8} 0.593} \\ 
\cline{2-8} \multicolumn{1}{|l|}{} &HyAB \cite{abasi2020distance} & \multicolumn{1}{c|}{\cellcolor[HTML]{fff8d6} 0.428} & \multicolumn{1}{c|}{\cellcolor[HTML]{ffd8b8} 0.447} & \multicolumn{1}{c|}{\cellcolor[HTML]{ffd8b8} 0.472} & \multicolumn{1}{c|}{\cellcolor[HTML]{f1f1f1} -} & \multicolumn{1}{c|}{\cellcolor[HTML]{cbe1cd} 0.329} & \multicolumn{1}{c|}{\cellcolor[HTML]{fdddd8} 0.593} \\ 
\cline{2-8} \multicolumn{1}{|l|}{} &ICtCp \cite{ITU-R_2124} & \multicolumn{1}{c|}{\cellcolor[HTML]{fff8d6} 0.427} & \multicolumn{1}{c|}{\cellcolor[HTML]{ffd8b8} 0.446} & \multicolumn{1}{c|}{\cellcolor[HTML]{fdddd8} 0.467} & \multicolumn{1}{c|}{\cellcolor[HTML]{f1f1f1} -} & \multicolumn{1}{c|}{\cellcolor[HTML]{cbe1cd} 0.329} & \multicolumn{1}{c|}{\cellcolor[HTML]{ffd8b8} 0.426} \\ 
\cline{2-8} \multicolumn{1}{|l|}{} &MS-SWD \cite{he2024ms-swd} & \multicolumn{1}{c|}{\cellcolor[HTML]{cbe1cd} 0.654} & \multicolumn{1}{c|}{\cellcolor[HTML]{ffd8b8} 0.44} & \multicolumn{1}{c|}{\cellcolor[HTML]{cbe1cd} 0.689} & \multicolumn{1}{c|}{\cellcolor[HTML]{f1f1f1} -} & \multicolumn{1}{c|}{\cellcolor[HTML]{fff8d6} 0.401} & \multicolumn{1}{c|}{\cellcolor[HTML]{fdddd8} 0.685} \\ 
\hline \hline
\multicolumn{1}{|l|}{\multirow{7}{*}{ \rotatebox[origin=c]{90}{ \begin{tabular}[c]{@{}c@{}}Deep-learning\\based metrics \end{tabular}}}} &WaDIQaM \cite{bosse2017deep} & \multicolumn{1}{c|}{\cellcolor[HTML]{fdddd8} 0.16} & \multicolumn{1}{c|}{\cellcolor[HTML]{fdddd8} 0.159} & \multicolumn{1}{c|}{\cellcolor[HTML]{fdddd8} 0.037} & \multicolumn{1}{c|}{\cellcolor[HTML]{f1f1f1} -} & \multicolumn{1}{c|}{\cellcolor[HTML]{f1f1f1} -} & \multicolumn{1}{c|}{\cellcolor[HTML]{ffd8b8} 0.393} \\ 
\cline{2-8} \multicolumn{1}{|l|}{} &LPIPS-Alex \cite{zhang2018perceptual} & \multicolumn{1}{c|}{\cellcolor[HTML]{ffd8b8} 0.393} & \multicolumn{1}{c|}{\cellcolor[HTML]{cbe1cd} 0.879} & \multicolumn{1}{c|}{\cellcolor[HTML]{cbe1cd} 0.839} & \multicolumn{1}{c|}{\cellcolor[HTML]{f1f1f1} -} & \multicolumn{1}{c|}{\cellcolor[HTML]{fff8d6} 0.408} & \multicolumn{1}{c|}{\cellcolor[HTML]{ffd8b8} 0.457} \\ 
\cline{2-8} \multicolumn{1}{|l|}{} &LPIPS-VGG \cite{zhang2018perceptual} & \multicolumn{1}{c|}{\cellcolor[HTML]{ffd8b8} 0.291} & \multicolumn{1}{c|}{\cellcolor[HTML]{fff8d6} 0.616} & \multicolumn{1}{c|}{\cellcolor[HTML]{cbe1cd} 0.903} & \multicolumn{1}{c|}{\cellcolor[HTML]{f1f1f1} -} & \multicolumn{1}{c|}{\cellcolor[HTML]{ffd8b8} 0.562} & \multicolumn{1}{c|}{\cellcolor[HTML]{fdddd8} 0.512} \\ 
\cline{2-8} \multicolumn{1}{|l|}{} &ST-LPIPS-Alex \cite{ghildyal2022shift} & \multicolumn{1}{c|}{\cellcolor[HTML]{fff8d6} 0.396} & \multicolumn{1}{c|}{\cellcolor[HTML]{ffd8b8} 0.495} & \multicolumn{1}{c|}{\cellcolor[HTML]{fdddd8} 0.408} & \multicolumn{1}{c|}{\cellcolor[HTML]{f1f1f1} -} & \multicolumn{1}{c|}{\cellcolor[HTML]{ffd8b8} 0.499} & \multicolumn{1}{c|}{\cellcolor[HTML]{fdddd8} 0.652} \\ 
\cline{2-8} \multicolumn{1}{|l|}{} &DISTS \cite{ding2020image} & \multicolumn{1}{c|}{\cellcolor[HTML]{ffd8b8} 0.324} & \multicolumn{1}{c|}{\cellcolor[HTML]{fff8d6} 0.561} & \multicolumn{1}{c|}{\cellcolor[HTML]{cbe1cd} 0.798} & \multicolumn{1}{c|}{\cellcolor[HTML]{f1f1f1} -} & \multicolumn{1}{c|}{\cellcolor[HTML]{fff8d6} 0.454} & \multicolumn{1}{c|}{\cellcolor[HTML]{cbe1cd} 0.249} \\ 
\cline{2-8} \multicolumn{1}{|l|}{} &AHIQ \cite{lao2022attentions} & \multicolumn{1}{c|}{\cellcolor[HTML]{ffd8b8} 0.358} & \multicolumn{1}{c|}{\cellcolor[HTML]{ffd8b8} 0.399} & \multicolumn{1}{c|}{\cellcolor[HTML]{fff8d6} 0.552} & \multicolumn{1}{c|}{\cellcolor[HTML]{f1f1f1} -} & \multicolumn{1}{c|}{\cellcolor[HTML]{f1f1f1} -} & \multicolumn{1}{c|}{\cellcolor[HTML]{fdddd8} 0.508} \\ 
\cline{2-8} \multicolumn{1}{|l|}{} &TOPIQ \cite{chen2024topiq} & \multicolumn{1}{c|}{\cellcolor[HTML]{ffd8b8} 0.309} & \multicolumn{1}{c|}{\cellcolor[HTML]{fdddd8} 0.101} & \multicolumn{1}{c|}{\cellcolor[HTML]{fdddd8} 0.096} & \multicolumn{1}{c|}{\cellcolor[HTML]{f1f1f1} -} & \multicolumn{1}{c|}{\cellcolor[HTML]{ffd8b8} 0.542} & \multicolumn{1}{c|}{\cellcolor[HTML]{fff8d6} 0.324} \\ 
\hline \hline
\multicolumn{1}{|l|}{\multirow{3}{*}{ \rotatebox[origin=c]{90}{ \begin{tabular}[c]{@{}c@{}}Video\\metrics \end{tabular}}}} &VMAF v0.6.1 \cite{li2018vmaf} & \multicolumn{1}{c|}{\cellcolor[HTML]{fff8d6} 0.577} & \multicolumn{1}{c|}{\cellcolor[HTML]{fff8d6} 0.549} & \multicolumn{1}{c|}{\cellcolor[HTML]{ffd8b8} 0.494} & \multicolumn{1}{c|}{\cellcolor[HTML]{fdddd8} 0.212} & \multicolumn{1}{c|}{\cellcolor[HTML]{fff8d6} 0.373} & \multicolumn{1}{c|}{\cellcolor[HTML]{f1f1f1} -} \\ 
\cline{2-8} \multicolumn{1}{|l|}{} &SpeedQA \cite{bampis2017speed} & \multicolumn{1}{c|}{\cellcolor[HTML]{ffd8b8} 0.285} & \multicolumn{1}{c|}{\cellcolor[HTML]{fdddd8} 0.111} & \multicolumn{1}{c|}{\cellcolor[HTML]{fff8d6} 0.556} & \multicolumn{1}{c|}{\cellcolor[HTML]{fdddd8} 0.182} & \multicolumn{1}{c|}{\cellcolor[HTML]{fdddd8} 1.775} & \multicolumn{1}{c|}{\cellcolor[HTML]{f1f1f1} -} \\ 
\cline{2-8} \multicolumn{1}{|l|}{} &FUNQUE \cite{venkataramanan2022funque} & \multicolumn{1}{c|}{\cellcolor[HTML]{fff8d6} 0.434} & \multicolumn{1}{c|}{\cellcolor[HTML]{cbe1cd} 0.714} & \multicolumn{1}{c|}{\cellcolor[HTML]{cbe1cd} 0.753} & \multicolumn{1}{c|}{\cellcolor[HTML]{fdddd8} 0.068} & \multicolumn{1}{c|}{\cellcolor[HTML]{ffd8b8} 0.52} & \multicolumn{1}{c|}{\cellcolor[HTML]{f1f1f1} -} \\ 
\hline \hline
\multicolumn{1}{|l|}{\multirow{6}{*}{ \rotatebox[origin=c]{90}{ \begin{tabular}[c]{@{}c@{}}Visual\\difference\\predictors \end{tabular}}}} &MAD \cite{larson2010most} & \multicolumn{1}{c|}{\cellcolor[HTML]{fdddd8} 0.125} & \multicolumn{1}{c|}{\cellcolor[HTML]{fff8d6} 0.528} & \multicolumn{1}{c|}{\cellcolor[HTML]{fdddd8} 0.42} & \multicolumn{1}{c|}{\cellcolor[HTML]{f1f1f1} -} & \multicolumn{1}{c|}{\cellcolor[HTML]{fdddd8} 0.861} & \multicolumn{1}{c|}{\cellcolor[HTML]{f1f1f1} -} \\ 
\cline{2-8} \multicolumn{1}{|l|}{} &HDR-VDP-3 \cite{mantiuk2023hdr} & \multicolumn{1}{c|}{\cellcolor[HTML]{cbe1cd} 0.695} & \multicolumn{1}{c|}{\cellcolor[HTML]{cbe1cd} 0.861} & \multicolumn{1}{c|}{\cellcolor[HTML]{cbe1cd} 0.897} & \multicolumn{1}{c|}{\cellcolor[HTML]{f1f1f1} -} & \multicolumn{1}{c|}{\cellcolor[HTML]{fff8d6} 0.372} & \multicolumn{1}{c|}{\cellcolor[HTML]{f1f1f1} -} \\ 
\cline{2-8} \multicolumn{1}{|l|}{} &FovVideoVDP \cite{mantiuk2021fovvideovdp} & \multicolumn{1}{c|}{\cellcolor[HTML]{fff8d6} 0.507} & \multicolumn{1}{c|}{\cellcolor[HTML]{cbe1cd} 0.934} & \multicolumn{1}{c|}{\cellcolor[HTML]{fff8d6} 0.538} & \multicolumn{1}{c|}{\cellcolor[HTML]{cbe1cd} 0.683} & \multicolumn{1}{c|}{\cellcolor[HTML]{ffd8b8} 0.573} & \multicolumn{1}{c|}{\cellcolor[HTML]{f1f1f1} -} \\ 
\cline{2-8} \multicolumn{1}{|l|}{} &ColorVideoVDP \cite{cvvdp} & \multicolumn{1}{c|}{\cellcolor[HTML]{cbe1cd} 0.89} & \multicolumn{1}{c|}{\cellcolor[HTML]{cbe1cd} 0.701} & \multicolumn{1}{c|}{\cellcolor[HTML]{fff8d6} 0.57} & \multicolumn{1}{c|}{\cellcolor[HTML]{cbe1cd} 0.518} & \multicolumn{1}{c|}{\cellcolor[HTML]{cbe1cd} 0.333} & \multicolumn{1}{c|}{\cellcolor[HTML]{cbe1cd} 0.027} \\ 
\cline{2-8} \multicolumn{1}{|l|}{} &CVVDP-ML-Saliency \cite{hammou2025colorvideovdp} & \multicolumn{1}{c|}{\cellcolor[HTML]{cbe1cd} 0.671} & \multicolumn{1}{c|}{\cellcolor[HTML]{cbe1cd} 0.728} & \multicolumn{1}{c|}{\cellcolor[HTML]{cbe1cd} 0.729} & \multicolumn{1}{c|}{\cellcolor[HTML]{ffd8b8} 0.293} & \multicolumn{1}{c|}{\cellcolor[HTML]{fdddd8} 0.599} & \multicolumn{1}{c|}{\cellcolor[HTML]{fff8d6} 0.343} \\ 
\cline{2-8} \multicolumn{1}{|l|}{} &CVVDP-ML-Transformer \cite{hammou2025colorvideovdp} & \multicolumn{1}{c|}{\cellcolor[HTML]{ffd8b8} 0.259} & \multicolumn{1}{c|}{\cellcolor[HTML]{fdddd8} 0.301} & \multicolumn{1}{c|}{\cellcolor[HTML]{fdddd8} 0.096} & \multicolumn{1}{c|}{\cellcolor[HTML]{fff8d6} 0.415} & \multicolumn{1}{c|}{\cellcolor[HTML]{f1f1f1} -} & \multicolumn{1}{c|}{\cellcolor[HTML]{fdddd8} 0.859}\\ 
\hline
\end{tabular}%
}
\end{table}

% As expected, the visual difference predictors that explicitly model low-level vision show the best alignment with the human data. LPIPS variants and DISTS stand out as those that model contrast masking well. MS-SWD models the detection of both chromatic and achromatic patterns well. Only FovVideoVDP and ColorVideoVDP model flicker. The RMSE values for contrast matching do not explain the performance well as they indicate that the metrics without spatial processing (e.g., PSNR-Y, CIEDE2000) perform the best. This is because those metrics have flat contrast response across all frequencies, which happen to capture contrast constancy at high contrast values --- see row (h) of \figref{metrics_results_main_fig}.

% \subsection{Contrast detection}
% \label{sec:contrast_detection}

\paragraph{Contrast detection}

One of the fundamental and well-studied visual characteristics is the eye's ability to detect near-threshold contrast, known as contrast detection, and modeled by the contrast sensitivity function (CSF). The CSF explains the smallest contrast that an average observer’s eye can detect on a uniform background across spatial and temporal frequencies, color, luminance, and stimulus area \cite{ashraf2024castlecsf}. While we evaluated metrics across all these dimensions, we focus here on spatial frequency, as it provides the most insight.
The contrast detection threshold for \emph{achromatic} patterns follows a band-pass function across spatial frequencies, shown as a red line in \figref{metrics_results_main_fig}-(a). The visual system exhibits the highest sensitivity (smallest contrast threshold) at intermediate frequencies between 2 and 4 cycles per visual degree (cpd) and a decrease in sensitivity at low and high frequencies. %The drop in sensitivity at low spatial frequencies is attributed to the lateral inhibition mechanism \cite{barten1999contrast} and to the optical limitations of the human eye \cite{campbell1965optical} at high frequencies. 

The contour plots in row (a) of \figref{metrics_results_main_fig}, together with the alignment scores reported in \tableref{metrics_results}, indicate that only a small subset of the evaluated metrics can reliably predict contrast sensitivity across spatial frequencies. 
Metrics such as PSNR, or color difference measures, such as CIEDE2000, have the same response across spatial frequencies. This is expected, as such simple per-pixel measures cannot estimate the effect of spatial frequencies. Among these metrics, MS-SWD (a color difference measure) is the only exception that achieves a good overall alignment with the CSF. Nevertheless, its response remains biased toward a low-pass characteristic. 
Metrics such as sCIELab and FLIP, rely on a simplified low-pass approximation of the CSF. As a result, they exhibit excessive sensitivity to low-frequency distortions.
In contrast, SSIM demonstrates a predominantly high-pass response, with peak sensitivity at high spatial frequencies. This behavior is inconsistent with the CSF and suggests that SSIM overemphasizes fine-scale distortions while underrepresenting mid-frequency distortions that are perceptually most salient. This limitation is partially mitigated in MS-SSIM, whose multi-scale formulation yields an effective band-pass response that more closely resembles the CSF.
Deep learning–based metrics such as WaDIQaM, and AHIQ exhibit erratic responses that do not align with the CSF. This behavior likely reflects their instability in accounting for distortions at different frequencies. DISTS, TOPIQ, LPIPS, and ST-LPIPS show clearer band-pass characteristics; however, their peak sensitivities occur at spatial frequencies that differ from those predicted by the CSF.
The closest agreement with the CSF is observed for ColorVideoVDP. This result is expected, as it is explicitly built upon the castleCSF, which is also used as the reference human sensitivity data in our analysis. However, its variant ColorVideoVDP-ML shows erratic responses, meaning that even when built upon CSF, its machine component impacts its performance in such a task.

\paragraph{Contrast masking}

Contrast detection, discussed in the previous section, applies to the case when a detected pattern (e.g., distortion) is shown on a uniform background. Such a case, however, is rare in natural images, as most distortions are mixed with image content, and the visibility of a distortion is masked by textures and patterns present in the image. This phenomenon is known as \emph{contrast masking} \cite{foley1994human}. %in the human vision literature and was found to be the strongest when the masking pattern has a similar spatial frequency and orientation as the detected pattern \cite{foley1994human}. 
Contrast masking explains how large the contrast of a pattern must be to be detected in the presence of a masker of a given contrast. The two examples of such masking characteristics are shown as red lines in rows (b) (data from \cite{foley1994human}) and (c) (data from \cite{gegenfurtner1992contrast}) of \figref{metrics_results_main_fig}. The test contrast required for detection increases once the masker's contrast is above the detection threshold. The difference between rows (b) and (c) is that in row (b) the masker phase is coherent with the test stimulus (see \figref{testing_methodology}-(b)) that is causing a facilitation (a ``dipper'' effect) of detection near the detection threshold, while in (c) the masker phase is incoherent with that of the test stimulus, which causes the facilitation to disappear. 
The metrics' performance on the contrast masking test can be found \tableref{metrics_results} and visualized in rows (b) and (c) of \figref{metrics_results_main_fig}. %The results reveal substantial variation in how different metrics capture masking behavior as a function of background contrast.

Several metrics, including PSNR-Y, SSIM, FLIP, and color difference measures, show little sensitivity to contrast masking, with discrimination thresholds remaining largely invariant to the masker contrast. FUNQUE, HDR-VDP-3, FSIMc, and VSI exhibit a straight increase of the discrimination threshold, reflecting a basic form of masking but failing to reproduce the reduced masking observed at low contrast levels below the detection threshold.
In contrast, FovVideoVDP, ColorVideoVDP, MS-SSIM, DSS, LPIPS, and DISTS demonstrate more nuanced behavior, capturing both the increase in masking at higher contrasts and a reduced masking at low contrasts. 
Interestingly, metrics derived from deep convolutional features, particularly those based on VGG and AlexNet (LPIPS, DISTS), exhibit a distinct dip in the masking contours (phase coherent test) consistent with contrast facilitation, a behavior not observed in most evaluated metrics. However, this dip occurs at contrast levels higher than those reported in psychophysical studies. 
Moreover, these metrics achieve the strongest performance in the phase-incoherent masking test, outperforming all other metrics. This suggests that deep feature representations generalize well to detecting contrast masking effects, particularly when the masker phase is incoherent to that of the masked stimulus, which resembles more closely the structural variability of natural images. In contrast, deep learning–based metrics relying on transformer architectures, such as AHIQ and TOPIQ, capture the presence of contrast masking at supra-threshold contrast levels. However, at low contrasts near the detection threshold, these metrics exhibit an anomalous decrease in predicted quality. A likely explanation is that transformer-based architectures lack sufficient sensitivity to subtle contrast variations and therefore mischaracterize masking effects at low contrast. 
Surprisingly, VMAF, despite strong performance on several quality benchmarks, exhibits contrast masking behavior, including a facilitation dip, only at supra-threshold contrasts. This suggests that its design and training primarily emphasize clearly visible distortions.

\paragraph{Flicker detection}

A flicker (rapid change of contrast/luminance over time) is a salient artifact that a video quality metric should be able to detect. %Flicker detection in human vision depends on temporal frequency, luminance, size of a flickering pattern, and the position in the visual field (eccentricity) \cite{Miller_Flicker_2023}, but here we focus only on the temporal frequency. 
The human sensitivity to flicker has a band-pass characteristic with a peak at 8\,Hz (see the red line in row (d) in \figref{metrics_results_main_fig}). When the critical flicker fusion frequency (CFF) is reached, the flicker is no longer perceived (is fused into a steady field due to the persistence of vision) \cite{cai2024elatcsf}.

Video metrics performance on the flicker test is reported in \tableref{metrics_results} and visualized in row (d) of \figref{metrics_results_main_fig}. Although multiple video quality metrics claim to model temporal distortions, only FovVideoVDP and ColorVideoVDP were able to predict the sensitivity to flicker. A part of the reason is that video quality metrics rarely consider more than a couple of frames when evaluating temporal distortions. Such a small filter is unable to differentiate between different temporal frequencies. ColorVideoVDP-ML variants show a similar performance to that of ColorVideoVDP; however, with some erratic behavior at supra-threshold contrasts.

% \subsection{Supra-threshold contrast matching}
% \label{sec:contrast_constancy}

\paragraph{Supra-threshold contrast matching across spatial frequencies} Many quality metrics incorporate the models of CSF and ensure band-pass (or low-pass) response across the spatial frequencies. However, the CSF predicts the visibility of only near-threshold contrast. Georgeson and Sullivan \cite{georgeson1975contrast} showed that for large contrasts that are much above the detection threshold, there is much less variation in the magnitude of perceived contrast across the spatial frequencies. Their contrast-matching data is shown as dashed lines in row (e) of \figref{metrics_results_main_fig}. The lines have a U-shape for small contrast, suggesting that both low and high frequencies need a higher contrast to match the perceived magnitude of medium frequencies. However, the dashed lines are almost flat at high contrast, indicating no difference in contrast perception across spatial frequencies. This phenomenon is known as \emph{contrast constancy}.

The results of contrast matching for each metric are reported in row (e) \figref{metrics_results_main_fig} as continuous lines, with different colors assigned to different contrast magnitudes. Note the missing segments of the line mean that contrast could not be matched to the reference. 
None of the quality metrics we tested could predict contrast constancy --- the ``flattening'' of the perceived contrast at large contrast values. In terms of contrast matching results, the quality metrics can be split into three groups. The first group is the metrics that do not contain any spatial processing, such as PSNR-Y or CIEDE2000. Those metrics maintain contrast constancy at large contrast but obviously cannot model the reduced perceived magnitude of small and medium contrast (their lack the U-shape in row (e) of \figref{metrics_results_main_fig}). The second group is the metrics that explicitly employ a contrast sensitivity function as a low-pass filter (sCIELab or FLIP) or a band-pass filter (ColorVideoVDP, HDR-VDP-3), or employ a frequency decomposition that modulates sensitivity with frequencies (MS-SSIM, DSS, HaarPSI, NLPD). For all metrics in this group, the lines of matching contrast are parallel to each other and show little flattening for high contrast. ColorVideoVDP and HDR-VDP-3 predict the CSF peak at lower frequencies and, therefore, show higher misalignment with the data of Georgeson and Sullivan. The final group is the metrics that fail to maintain contrast response across frequencies and show erratic behavior, such as VIFp, FSIMc, VSI, TOPIQ, and LPIPS-VGG. The metrics behavior in this test is similar to that of the contrast detection test.

\paragraph{Contrast matching across color directions}
% We want metrics to correctly estimate the magnitude of perceived contrast not only across spatial frequencies but also across different color directions. To test for that, we rely on the data of Switkes and Crognale \cite{switkes1999comparison}, who matched contrast along different contrast modulation directions in a color space, as shown in column (d) of \figref{testing_methodology}. The data allowed us to generate square-wave gratings in the achromatic (grayscale), red-green, and yellow-violet color directions so that the perceived magnitude of contrast for each is matched. In row (i) of \figref{metrics_results_main_fig}, we plot the metrics' response of each such triplet, connected by a line. If a metric correctly matches contrast across color directions, the response of the metric should be the same, and the lines should be horizontal. 

Perceptually accurate color quality metrics or color difference measures should be able to match contrast across the achromatic, red–green, and yellow–violet directions, reflecting the differing human sensitivities along these channels. The performance of the evaluated metrics on this test is reported in \tableref{metrics_results}, with representative plots shown in row (f) of \figref{metrics_results_main_fig}. The results indicate that several metrics do not balance perceived contrast magnitudes consistently. For instance, CIEDE2000 and HyAB are substantially more sensitive to achromatic than to chromatic contrast, whereas sCIELab, FLIP, MS-SWD, and ICtCp exhibit the opposite tendency.
It should be noted that metrics based on color difference formulas (e.g., CIEDE2000) were originally calibrated for small, near-threshold color differences, while our test includes large supra-threshold contrasts. Deep learning–based metrics, including LPIPS, TOPIQ, and DISTS, generally overemphasize chromatic differences. Overall, ColorVideoVDP achieved the closest match to balanced perceptual contrast, followed by FSIMc, which performs well at low contrasts, and DISTS, which again tends to overestimate chromatic effects.

% \section{Aggregate performance measures}

% In addition to contour plots, we also calculate summative performance measures of alignment score and RMSE. Those are identical as in \cite{cai2024elatcsf} and also explained in \suppsecref{performance-measures}. Such measures lack the interpretation and detail provided by the contour plots but give a high level overview and facilitate comparison across multiple metrics. 

% \tableref{metrics_results} shows the alignment scores (the higher, the better) for contrast detection and masking and RMSE (the lower, the better) for contrast matching. As expected, the visual difference predictors that explicitly model low-level vision show the best alignment with the human data. LPIPS variants and DISTS stand out as those that model contrast masking well. MS-SWD models the detection of both chromatic and achromatic patterns well. Only FovVideoVDP and ColorVideoVDP model flicker. The RMSE values for contrast matching do not explain the performance well as they indicate that the metrics without spatial processing (e.g., PSNR-Y, CIEDE2000) perform the best. This is because those metrics have flat contrast response across all frequencies, which happen to capture contrast constancy at high contrast values --- see row (h) of \figref{metrics_results_main_fig}.

\section{Conclusion}
\label{sec:conclusion}

The methodology proposed in this paper allows us to test whether quality metrics model the low-level characteristics of the visual system. This led us to multiple interesting and new observations about the performance of widely used quality metrics. For example, we found that many deep learning metrics excel in modeling contrast masking (e.g., LPIPS, DISTS) even though they have never been trained on such data. That popular SSIM overemphasized differences in higher frequencies, contrary to contrast detection and contrast matching psychophysical data. That VMAF, which is one of the most widely used video metrics, models the effect of contrast masking for only supra-threshold contrasts. Or that color difference formulas (CIEDE2000, HyAB) underestimate color differences for large (supra-threshold) contrast levels. We hope that our methodology and software will help in scrutinizing the performance of existing and newly proposed metrics. 

% It is important to state that a good performance in our low-level human vision tests may not be a required or sufficient condition for a metric to perform well across the applications. For example, SSIM is arguably better at capturing perceived quality than PSNR, but \tableref{metrics_results} shows that for all tests except contrast masking, PSNR-Y is better aligned with the visual system. Instead, we put an argument that a metric that performs well in our low-level vision tests has a better chance to generalize to new and unseen types of distortions. Our method is also limited to full-reference metrics that quantify the fidelity to the reference. Our method cannot be used with no-reference metrics that assess other aspects of quality, such as aesthetics or how well the distribution of results aligns with the distribution of natural images (e.g., FID metric). 

\bibliographystyle{IEEEtran}
\bibliography{ref}

\end{document}